# Mössbauer studies of spin- and charge-modulations in BaFe$_2$(As$_{1-x}$P$_x$)$_2$

K. Komędera[1], J. Gatlik[1], A. Błachowski[1*], J. Żukrowski[2], T. J. Sato[3],
and
D. Legut[4], U. D. Wdowik[5]

[1]Mössbauer Spectroscopy Laboratory, Institute of Physics, Pedagogical University, Kraków, Poland
[2]Academic Centre for Materials and Nanotechnology, AGH University of Science and Technology, Kraków, Poland
[3]Institute of Multidisciplinary Research for Advanced Materials, Tohoku University, Sendai, Japan
[4]IT4Innovations , VSB-Technical University of Ostrava, Ostrava, Czech Republic
[5]Institut of Technology, Pedagogical University, Kraków, Poland

[*]Corresponding author: sfblacho@cyf-kr.edu.pl

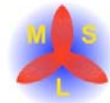

**Abstract**

The BaFe$_2$(As$_{1-x}$P$_x$)$_2$ compounds with x = 0 (parent), x = 0.10 (under-doped), x = 0.31, 0.33, 0.53 (superconductors with $T_c$ = 27.3 K, 27.6 K, 13.9 K, respectively) and x = 0.70, 0.77 (over-doped) have been investigated versus temperature using $^{57}$Fe Mössbauer spectroscopy. Special attention was paid to regions of the spin-density-wave (SDW) antiferromagnetic order, spin-nematic phase, and superconducting transition. The BaFe$_2$(As$_{0.90}$P$_{0.10}$)$_2$ compound exhibits a reduced amplitude of SDW as compared to the parent compound and preserved universality class of two-dimensional magnetic planes with one-dimensional spins. The spin-nematic phase region for x = 0.10 is characterized by an incoherent magnetic order. BaFe$_2$(As$_{0.69}$P$_{0.31}$)$_2$ shows coexistence of a weak magnetic order and superconductivity due to the vicinity of the quantum critical point. The charge density modulations in the BaFe$_2$(As$_{0.67}$P$_{0.33}$)$_2$ and BaFe$_2$(As$_{0.47}$P$_{0.53}$)$_2$ superconductors are perturbed near $T_c$. Pronounced hump of the average quadrupole splitting across superconducting transition is observed for the system with x = 0.33. The phosphorus substitution increases the Debye temperature of the BaFe$_2$(As$_{1-x}$P$_x$)$_2$ compound. Moreover, experimental electron charge densities at Fe nuclei in this material conclusively show that it should be recognized as a hole-doped system. The measured Mössbauer spectral shift and spectral area are not affected by transition to the superconducting state. This indicates that neither the average electron density at Fe nuclei nor the dynamical properties of the Fe-sublattice in BaFe$_2$(As$_{1-x}$P$_x$)$_2$ are sensitive to the superconducting transition. Theoretical calculations of hyperfine parameters determining the patterns of Mössbauer spectra of BaFe$_2$(As$_{1-x}$P$_x$)$_2$ with x = 0, 0.31, 0.5, and 1.0 are performed within the framework of the density functional theory. These simulations provide an insight into changes of the immediate neighborhood experienced by Fe atoms upon the P-for-As substitution as well as enable us to explore influence of P-doping on the electron density, electric field gradient, and hyperfine field at Fe nuclei in the BaFe$_2$(As$_{1-x}$P$_x$)$_2$ system.

## 1. Introduction

BaFe$_2$As$_2$ is a metallic parent compound of the iron-based superconductors belonging to '122' family. A tetragonal-to-orthorhombic structural transition at 139 K which is accompanied by development of the antiferromagnetic order of itinerant iron moments is observed upon cooling. The magnetic order has the spin-density-wave (SDW) character with small iron moments (< 1 µ$_B$) aligned antiferromagnetically along the orthorhombic *a* axis [1-3]. The SDW-type itinerant antiferromagnetic order results from an instability due to the Fermi surface nesting between electron and hole pockets. Superconductivity can be induced by chemical substitution (doping) or hydrostatic pressure. The BaFe$_2$(As$_{1-x}$P$_x$)$_2$ compounds represent 'Ba-122' iron-pnictide superconductors with superconductivity induced by nominally "isovalent" chemical doping of the BaFe$_2$As$_2$ parent compound (e.g., the partial substitution of As by P) with wide superconducting dome within range 0.2 < x < 0.7 [4]. The highest critical temperature $T_c$ = 31 K appears for x ≈ 0.3. The P-for-As substitution introduces internal strain or distortion, i.e., chemical pressure because P anion is smaller than As anion. In general, the Fe-based superconductors show high pressure dependence of $T_c$ reaching +10 K/GPa for under-doped BaFe$_2$(As$_{0.8}$P$_{0.2}$)$_2$ and -3 K/GPa for a superconductor with x = 0.35 [5, 6]. The phosphorus substitution simultaneously suppresses the orthorhombic distortion and the SDW magnetic order with complete disappearance of both transitions at about x ≈ 0.3 [7, 8]. The *c*-axis compression with reduction of the average distances of pnictogen anions above the iron plane, from h$_{As}$ = 1.36 Å for x = 0 to h$_{As/P}$ = 1.28 Å for x = 0.3, was found [9]. The same substitution level x for suppression of the static long-range magnetic order and for emergence of the highest $T_c$ indicates the interplay between magnetism and superconductivity in these compounds. This suggests the existence of the quantum critical point (QCP) near x = 0.3. The complex magnetic and electronic phase diagrams are common for the Fe-based superconductors and may result in the appearance of QCP [10-14]. It is believed that quantum-critical fluctuations originating from QCP are closely related with superconductivity via unconventional pairing mechanism. Most microscopic theories describing the Fe-based superconductors focus on the role of spin- [15, 16], orbital- [17], or nematic- [18, 19] fluctuations in the electron pairing mechanism [20]. On the other hand, both spin-density-wave and charge-density-wave (CDW) instabilities may compete with other possible orderings, and their interplay complicates the phase diagram of iron-pnictide systems.

It is known that $^{57}$Fe Mossbauer spectroscopy is sensitive to the shape and amplitude of SDW and CDW [21-24]. Hence, it is interesting to investigate changes of SDW in under-doped compounds and explore possible existence of the magnetic order near optimal-doping. Electronic charge modulations in the iron-based superconductors were observed by $^{57}$Fe Mossbauer spectroscopy as the *s*-electrons charge density modulation (CDW) and non-*s* electron (mainly *d*-electrons) electric field gradient wave (EFGW) [23, 24]. The CDW is a spatial modulation of charge (electron) density and it can be approximated by time independent standing plane wave with spatial period, which is quite often incommensurate with the lattice period. The *s*-electrons of CDW affect the nuclear states of resonant nuclei leading to distribution of the isomer shifts. Similarly to the modulation of density of electrons with angular momentum greater than zero, modulation of the electric field gradient (EFG) can be expected as well. This effect is usually superimposed on the constant EFG term arising from local symmetry lower than cubic or the EFG distribution associated with chemical substitution. These modulations are sensitive to the superconducting transition, i.e., they are perturbed upon transition to superconducting state and they become recovered upon full development of the superconducting gap. Such effects were observed for Ba$_{0.6}$K$_{0.4}$Fe$_2$As$_2$ [23] with superconductivity induced by hole-doping (partial substitution of Ba by K) and

SmFeAsO$_{0.91}$F$_{0.09}$ [24] with superconductivity induced by electron-doping (partial substitution of O by F). In close vicinity of $T_c$, the effects of CDW enhancement and EFGW suppression are observed in SmFeAsO$_{0.91}$F$_{0.09}$, whereas in Ba$_{0.6}$K$_{0.4}$Fe$_2$As$_2$ CDW and EFGW are respectively suppressed and enhanced. The differences between SmFeAsO$_{0.91}$F$_{0.09}$ and Ba$_{0.6}$K$_{0.4}$Fe$_2$As$_2$ follow from a different type of doping (electron and hole) required to realize superconductivity in these compounds. The 'Ba-122' superconductors with electron-doping, such as Ba(Fe$_{1-x}$T$_x$)$_2$As$_2$ have almost the same $^{57}$Fe Mössbauer spectral shift as the parent BaFe$_2$As$_2$ compound (see Discussion section for more details), and therefore they can be less useful for searching a similar phenomenon. Hence, it is interesting to look for similar effects in compounds with superconductivity induced by the isovalent chemical doping. To investigate the spin- and charge-modulations, the BaFe$_2$(As$_{1-x}$P$_x$)$_2$ samples with x = 0.10 (under-doped), x = 0.31, 0.33, 0.53 (superconductors) and x = 0.70, 0.77 (over-doped) have been measured as a function of temperature by $^{57}$Fe Mössbauer spectroscopy. The superconducting compound with x = 0.32 has been investigated previously by A. Sklyarova *et al.* [25]. The results of our Mössbauer spectroscopy experiments are supported by the *ab initio* calculated hyperfine parameters for the BaFe$_2$(As$_{1-x}$P$_x$)$_2$ systems with x = 0, 0.31, 0.50, and 1. These theoretical studies, which are based on the state-of-the-art density functional theory (DFT) allowed us to clarify a complexity of the measured spectra and relate it to the changes in local environments of the resonant $^{57}$Fe nuclei originating from nominally isovalent doping.

## 2. Experimental

The single crystals of the BaFe$_2$(As$_{1-x}$P$_x$)$_2$ compounds were grown by the flux method using mechanically alloyed precursors, as described in Ref. [26]. The XRD analysis confirmed that the BaFe$_2$(As$_{1-x}$P$_x$)$_2$ crystals are single-phase samples. The compositions of samples and the substitution level x were determined by the EDX analysis [26]. The critical temperature $T_c$ of the investigated superconductors was determined from the magnetic susceptibility drop (midpoint), while the SDW magnetic ordering temperature $T_{SDW}$ for the under-doped sample was obtained from temperature of the inflection point in the resistivity [26].

$^{57}$Fe Mössbauer spectroscopy measurements were performed in transmission geometry using the RENON MsAa-4 spectrometer operated in round-corner triangular mode and equipped with the LND Kr-filled proportional detector. The He-Ne laser-based Michelson–Morley interferometer was used to calibrate a velocity scale. A commercial $^{57}$Co(Rh) source made by RITVERC GmbH was kept at room temperature. The source linewidth $\Gamma_s = 0.106(5)$ mm/s and the effective source recoilless fraction were derived from fit of the Mössbauer spectrum of 10-μm-thick α-Fe foil. The Mössbauer absorbers were prepared in powder form using 35 mg of BaFe$_2$(As$_{1-x}$P$_x$)$_2$ compounds mixed with a B$_4$C carrier. The absorber thickness (surface density) amounted to 17 mg/cm$^2$ of the investigated material. The Janis Research Inc. SVT-400 cryostat was used to maintain temperature of absorbers in a range 4.2 – 300 K. Uninterrupted series of measurements have been carried out in the temperature ranges of 4.2–75 K and 80–300 K. Liquid helium and liquid nitrogen were used as cooling media for two non-interrupted series of measurements on each sample. The geometry, count-rate, and single channel analyzer window borders were kept constant during measurements. The recorded data were processed by the MOSGRAF software suite. The average center shifts <δ> of the measured Mössbauer spectra are given with respect to the center shift of the room temperature α-Fe.

*2.1 DFT calculation methods*

To gain a stronger understanding of the underlying structural and electronic features that give rise to complex Mössbauer spectra of the examined BaFe$_2$(As$_{1-x}$P$_x$)$_2$ compounds, the DFT calculations of the Mössbauer hyperfine parameters were performed. The compositions at x = 0.0, 0.31, 0.5, 1.0 were modeled by the 1×1×1 (x = 0.0, 0.5, 1.0) and 2×2×1 (x = 0.31) supercells obtained from the ideal BaFe$_2$As$_2$ structure of *I*4/*mmm* (No. 139) space group, in which Ba atoms occupy 2*a*(0,0,0) positions, Fe atoms are located at 4*d*(0,½,¼) positions, and As/P atoms reside at 4*e*(0,0,*z*) sites. To investigate the effect of local environment on the hyperfine parameters at Fe nuclei due to substitution of As atoms with P atoms, various configurations including four As atoms (4As), three As and one P atoms (3As1P), two As and two P atoms (2As2P), one As and three P atoms (1As3P), and four P atoms (4P) as the nearest-neighbors of Fe atoms have been simulated. Calculations have been performed within the spin-polarized density functional theory (DFT) method using the plane-wave basis VASP code [27] and the full-potential (linearized) augmented plane-wave plus local orbitals [FP-(L)APW + lo] method, the latter implemented in the WIEN2K code [28]. The pseudopotential plane-wave method has been applied to optimize the structures, while the all-electron method has been applied to calculate hyperfine interactions at Fe nuclei.

*Calculations with the pseudopotential method.* Electron–ion interactions were described by the projector augmented wave (PAW) method [29]. The PAW pseudopotentials with reference configurations for valence electrons Ba(5s$^2$5p$^6$6s$^2$), Fe(3d$^7$4s$^1$), As(4s$^2$4p$^3$), and P(3s$^2$3p$^3$) were taken from VASP database. The gradient-corrected exchange-correlation functional with parametrization of Pedrew, Burke, and Ernzerhof (GGA-PBE) revised for solid systems (PBEsol) [30] together with a plane-wave expansion up to 400 eV were applied. Spin-polarized calculations with the type-I antiferromagnetic order (AF-I), in which the magnetic moments on Fe atoms are aligned within (100) layer and opposite to the moments of next (100) layer, resulted in the magnetic moments on Fe atoms less than 0.04 $\mu_B$. The Brillouin zones of the BaFe$_2$(As$_{1-x}$P$_x$)$_2$ structures were sampled with the Monkhorst-Pack *k*-point meshes of 12×12×4 (x = 0.0, 0.5, 1.0) and 6×6×4 (x = 0.31). The ideal structures with x = 0.0 and x = 1.0 were fully optimized. The resulting lattice constants and internal atomic coordinates for BaFe$_2$As$_2$ (*a* = 3.8845 Å, *c* = 12.4580 Å, *z* = 0.3471), and BaFe$_2$P$_2$ (*a* = 3.7623 Å, *c* = 12.2530 Å, *z* = 0.3395) remain in close correspondence to those reported previously [1, 31]. Structural relaxations of the BaFe$_2$(As$_{1-x}$P$_x$)$_2$ systems with x = 0.31 and x = 0.5 were performed for fixed volumes of the supercells followed from optimized value of BaFe$_2$As$_2$ lattice. Convergence criteria for the residual forces and total energies of 0.01 eV/Å and 0.01 meV were applied.

*Calculations with the full-potential method.* The scalar-relativistic FP-(L)APW method was applied. The wave functions have been expanded into spherical harmonics inside the nonoverlapping atomic spheres having the radii $R_{MT}$ and in the plane waves within the interstitial region. The $R_{MT}$ were set to 2.5, 2.2, 2.0, 1.85 a.u. for Ba, Fe, As, and P, respectively. The maximum *l* value for expansion of the wave functions into the spherical harmonics inside $R_{MT}$ spheres was set to $l_{max} = 10$, while for expansion of the wave functions within the interstitial region, the plane-wave cutoff parameter $K_{max} = 7/R^{min}_{MT}$ was applied. The charge density was Fourier-expanded up to $G_{max} = 12$ Ry$^{1/2}$. States lying more than 8 Ry below the Fermi level were treated as the core states. Alike in calculations with the pseudopotential method, the AF-I magnetic arrangement was taken into account and the correlation and exchange potentials were treated within the GGA-PBEsol approximation. The calculations have been carried out for the supercell volumes adopted from the pseudopotential method calculations. Only the atomic positions were relaxed within the FP-LAPW methodology with the force convergence of 0.01 mRy/a.u.. The Brillouin zone integrations were performed with similar Monkhorst-Pack *k*-point meshes as those used in pseudopotential calculations. All self-consistent calculations have been performed with the energy and charge convergence criteria better than $10^{-6}$ Ry. The isomer shift has been obtained from the calculated electron contact densities at resonant nucleus ρ in a given matrix and reference material $ρ_0$ as $δ = α(ρ–ρ_0)$, where α stands for the calibration constant characteristic for a particular nuclear transition. Our calculations are performed for the 14.41-keV transition in $^{57}$Fe, and the isomer shifts are given with respect to the metallic bcc α-Fe upon applying previously determined α = 0.291 (a.u.)$^3$·mm/s [32]. The iron nucleus has been approximated by homogeneously charged sphere with the radius $R_0$ = 4.897 fm. Further technical details of such calculations can be found in our earlier papers [33].

## 3. Results

The representative $^{57}$Fe Mössbauer spectra measured at 4.2 K, 80 K and 300 K for the BaFe$_2$As$_2$ parent compound and the BaFe$_2$(As$_{1-x}$P$_x$)$_2$ compounds are shown in **Fig. 1**. The room temperature spectra of the parent compound show a single narrow line, whereas the spectra of substituted samples exhibit an asymmetrically broadened pseudo-single line. Both singlet and pseudo-singlets follow from very small electric quadrupole hyperfine interaction, i.e., small EFG at Fe atoms occupying 4d Wyckoff positions in the tetragonal structure of the *I4/mmm* space group. One also notes that the principal axis of the EFG tensor lies along the crystallographic *c*-axis. The Mössbauer spectra of the BaFe$_2$(As$_{1-x}$P$_x$)$_2$ compounds are more complex and require fitting with the distribution of quadrupole-split doublets due to doping-induced disorder and perturbation of the Fe-As layers by P atoms which are almost randomly distributed.

The development of the SDW magnetic order with roughly rectangular shape is seen at 4.2 K for the parent and under-doped compounds. The shape of spin modulation indicates that the SDW order is almost indistinguishable from the simple antiferromagnetic order close to the ground state, as long as the Mössbauer spectroscopy is concerned.

A transmission integral approximation has been applied to fit Mössbauer spectra exhibiting magnetic hyperfine interaction. The absorption profile of the SDW spectral component was processed by applying a quasi-continuous distribution of the magnetic hyperfine field *B*. For the Fe magnetic moment being collinear with the hyperfine filed, the amplitude of SDW along the direction parallel to the SDW wave vector can be expressed as a series of odd harmonics [21, 34, 35]:

$$B(qx) = \sum_{n=1}^{N} h_{2n-1} \sin[(2n-1)qx], \tag{1}$$

where *B(qx)*, *q*, *x*, and *qx* denote respectively the magnetic hyperfine field arising from SDW, the wave number of SDW, relative position of Fe atom along the propagation direction of stationary SDW, and phase shift. The symbols $h_{2n-1}$ denote the amplitudes of subsequent harmonics. The index *N* enumerates maximum relevant harmonic. Amplitudes up to seven subsequent harmonics (*N* = 7) have been fitted to the spectral shape. The amplitude of the first harmonic is positive, as absolute phase shift between SDW and the crystal lattice is generally unobservable by the method used. The SDW shape changes from almost rectangular for parent-compound to quasi-triangular for doped-compounds, and hence *N* = 4 was found sufficient to describe spectra for the superconducting compound. The constant number of *N* harmonics was chosen in such a way as to obtain a good-quality fit of the experimental spectrum represented by the parameter $χ^2$ per degree of freedom of the order of 1.0. The argument *qx* satisfies the following condition: $0 ≤ qx ≤ 2π$ due to the periodicity of the SDW. Further details of the Mössbauer spectra evaluation within SDW model can be found in Ref. [21]. Because the average amplitude of SDW described by expression (1) equals zero, thus the root-mean-square amplitude of SDW expressed as $<B> = \sqrt{\langle B^2 \rangle} = \sqrt{\frac{1}{2}\sum_{n=1}^{N} h_{2n-1}^2}$ was used as the average magnetic hyperfine field of the SDW [35]. Here, the parameter $<B>$ is proportional to the value of the electronic magnetic moment per unit volume.

The under-doped $BaFe_2(As_{0.90}P_{0.10})_2$ sample presents strongly perturbed SDW magnetic order with $T_{SDW}$ = 106 K determined on the basis of anomaly in electrical resistivity [26]. At 4.2 K, the average hyperfine magnetic field of SDW $<B>$ = 4.11 T is significantly reduced in comparison with the value of the parent compound $BaFe_2As_2$ with $<B>$ = 5.28 T. The measured hyperfine magnetic field can be estimated as $<B> = \alpha\, \mu_{Fe}$, where $\mu_{Fe}$ is the on-site magnetic moment of iron atom and $\alpha$ denotes the constant, which is specific for a given compound [36]. To convert the hyperfine magnetic field to the iron magnetic moment, we used $\alpha$ = 6.1 T/$\mu_B$, which results from $<B>$ obtained for $BaFe_2As_2$ [21] and $\mu_{Fe}$ = 0.87 $\mu_B$ determined from the neutron-diffraction measurements [37]. Hence, the hyperfine magnetic field $<B>$ at 4.2 K for under-doped $BaFe_2(As_{0.90}P_{0.10})_2$ corresponds to $\mu_{Fe}$ = 0.67 $\mu_B$. Similarly, small values of $\mu_{Fe}$ and the respective $\alpha$ constants were determined from the neutron diffraction and Mössbauer studies of other '122' parent compounds of Fe-based superconductors. For example, $\mu_{Fe}$ = 0.80 $\mu_B$ ($\alpha$ = 11.2 T/$\mu_B$) for $CaFe_2As_2$ [38], $\mu_{Fe}$ = 0.94 $\mu_B$ ($\alpha$ = 9.5 T/$\mu_B$) for $SrFe_2As_2$ [39], and $\mu_{Fe}$ = 0.99 $\mu_B$ ($\alpha$ = 8.1 T/$\mu_B$) for $EuFe_2As_2$ [40] have been reported. Different scenarios were suggested to explain small values of iron magnetic moments. Most likely such small magnetic moments are related to a prominent itinerant character of iron spins.

The hyperfine field $<B>$ in $BaFe_2(As_{0.90}P_{0.10})_2$ rapidly decreases above $T_{SDW}$, but the residual magnetic order is observed up to 130 K, i.e., about 25 K above the temperature of the coherent SDW order (see **Fig. 2** and **Fig. 3**). Between $T_{SDW}$ and 130 K, the under-doped $BaFe_2(As_{0.90}P_{0.10})_2$ is found in the spin-nematic phase [7], which is characterized by electronic anisotropy in the crystallographic *a-b* plane with broken rotational symmetry, but preserved translational symmetry. The results of the Mössbauer spectroscopy suggest that the spin-nematic phase is a region of incoherent spin density wavelets that are typical for a critical region. Temperature evolution of the SDW hyperfine magnetic field $<B>(T)$, which is shown for $BaFe_2(As_{0.90}P_{0.10})_2$ in **Fig. 3**, was fitted within the model described in Ref. [21]. The static critical exponent of 0.13(1) and the coherent SDW order temperature of 110(1) K were obtained. We note that the critical exponent in the substituted system $BaFe_2(As_{0.90}P_{0.10})_2$ is the same as that in the parent compound [3, 21]. This indicates that the universality class (1, 2) is retained in $BaFe_2(As_{0.90}P_{0.10})_2$. It also shows that the electronic spin system with SDW obeys the Ising model (one-dimensional spin space) and has two dimensions in the configuration space (magnetized planes).

The $BaFe_2(As_{0.69}P_{0.31})_2$ superconductor with $T_c$ = 27.3 K shows traces of the magnetic order below 60 K, i.e., also in its superconducting state, see **Fig. 2** and **Fig. 3**. The hyperfine field $<B>$ (4.2 K) = 2.20 T corresponds to approximately 0.36 $\mu_B$ of iron moment. The coexistence of magnetism and superconductivity is probably due to vicinity of the quantum critical point reported for this composition [41]. Despite a small jump of about 0.1 T between 26 K and 28 K, which slightly exceeds the error of measured values, no change of the average magnetic field $<B>$ at the critical temperature is observed. The spectra above 60 K, where no magnetic dipole hyperfine interaction is detected, are fitted with the distribution of EFG.

The shapes of Mössbauer spectra without magnetic interaction are influenced by non-vanishing hyperfine electric quadrupole interaction arising from both the crystallographic structure of $BaFe_2(As_{1-x}P_x)_2$ compounds and distortions of the Fe-As layers caused by the inserted P atoms. The non-magnetic spectra are fitted with the average electric quadruple splitting $<\Delta>$ resulting from distribution of the quadrupole-split doublets determined by the

Hesse-Rübartsch method in the Lorentzian approximation [42, 43]. The distribution of quadrupole-split doublets originates from a random substitution of As atoms by P atoms which yields different configurations around iron atoms and finally leads to a perturbation of the local EFG. The spectral linewidth $\Gamma$ within the Hesse-Rübartsch distribution fitting method was kept constant and equal to the natural linewidth.

The Mössbauer spectra of $BaFe_2(As_{0.67}P_{0.33})_2$ with $T_c$ = 27.6 K and $BaFe_2(As_{0.47}P_{0.53})_2$ with $T_c$ =13.9 K display no magnetic dipole hyperfine interaction in the whole temperature range, and thus they are fitted using the distribution of EFG. The EFG distribution is associated with perturbation of the iron surrounding by the phosphorus atoms in the Fe-As layer. A relatively wide distribution range with the quadrupole splitting $\Delta$, reaching up to about 1.6 mm/s (or more) is also observed in other optimally doped Fe-based superconductors [23, 24]. It is related with the spatial modulation of EFG, which is called the electric field gradient wave (EFGW). Such large splitting $\Delta$ could be observed for highly covalent bonds of iron with the electron(s) located in one of the lobes of 3d 'atomic' state. This electronic configuration is consistent with the observed (and reported in the text below) spectral center shift of about 0.5 mm/s at the ground state. It was found that the average quadrupole splitting $<\Delta>$ varies at the critical temperature for x = 0.33, i.e., for the optimally doped superconductor with the highest $T_c$ in our studies, see **Fig. 2** and **Fig. 3**. Variation is characterized by an increase in $<\Delta>$ at $T_c$ probably due to opening of the superconducting gap and subsequent formation of the Cooper pairs. The similar effect has been previously observed for other Fe-based superconductors with perturbation of the spatial modulation of EFG resulting from the incommensurate modulation of the electron charge density at Fe nuclei [23, 24].

Despite some subtle change in the Mössbauer spectra shape of the $BaFe_2(As_{0.47}P_{0.53})_2$ superconductor near $T_c$ (see **Fig. 3**), no change in $<\Delta>$ at $T_c$ is detected when applying the Hesse-Rübartsch distribution fitting model. A tiny change in the spectrum shape can be visible only when the spectral linewidth $\Gamma$ of the Hesse-Rübartsch distribution is not a constant value. In such a case, the oscillations of the $\Gamma$ of about 0.04 mm/s across $T_c$ may be observed. It is known that a signature of CDW can be seen as the excess of Mössbauer spectral linewidth [23, 24]. Oscillations of the spectral linewidth $\Gamma$ due to sole variation of the isomer shift correspond to the oscillations of electron density at Fe nuclei [32] across formation of the superconducting state of about 0.15 el./(Bohr radius)$^3$. On the other hand, no change in the electron charge density modulations and spectral shapes was observed for over-doped compounds with x = 0.70 and 0.77 in the whole investigated temperature range.

## 4. Discussion

Temperature dependencies of the average center shift $<\delta>$, which are given with respect to α-Fe at room temperature, are shown in **Fig. 3**. At 4.2 K, the $<\delta>$ changes from 0.55 mm/s for parent compound to 0.45 mm/s for over-doped compound with x = 0.77, whereas at 300 K, the $<\delta>$ changes from 0.43 mm/s for parent compound to 0.33 mm/s for over-doped one. The value of this hyperfine parameter indicates that Fe atoms are in a low-spin Fe(II) electronic configuration. A decrease in $<\delta>$ with increasing substitution level x shows that phosphorus atoms increase the *s*-electrons charge density at Fe nuclei in $BaFe_2(As_{1-x}P_x)_2$ due to weakening of the shielding effect caused by *d*-electrons. Therefore, the replacement of As by P decreases the conducting *d*-electrons charge density and can be interpreted as hole-doping. Indeed, for the optimally doped $BaFe_2(As_{0.67}P_{0.33})_2$ superconductor the shift $<\delta>$ equals to 0.51 mm/s at 4.2 K and 0.39 mm/s at RT which is almost exactly the same as for optimally hole-doped $Ba_{0.6}K_{0.4}Fe_2As_2$ at the same temperatures [23]. Sizable number of holes in the hole Fermi surfaces induced by P-for-As substitution in BaFe$_2$(As$_{1-x}$P$_x$)$_2$ was also found by ARPES spectroscopy [6]. On the other hand, the <δ> for optimally electron-doped Ba-122 superconductors such as Ba(Fe$_{1-x}$T$_x$)$_2$As$_2$ (T = Co [35, 44], Ni [45], Rh [46]) amounts to 0.55 mm/s at the ground state, which is exactly the same as for BaFe$_2$As$_2$. This result raised the question "whither the injected electron(s) go?" [47]. The lack of increase in *d*-shell population observed in Mössbauer experiments [47] was accounted for by hybridization between *p*- and *d*-orbitals in Fe–As bonding as well as lattice contraction upon substitution which enhances the *sp* hybridization in As-As bonds and provides a path for the migration of additional electrons. It should also be taken into account that the level of substitution with the transition metal T, which is sufficient for optimal doping, usually remains low (much smaller than 0.1). The Ba(Fe$_{1-x}$T$_x$)$_2$As$_2$ system also offers an opportunity for isovalent doping with T = Ru. In the Ba(Fe$_{0.5}$Ru$_{0.5}$)$_2$As$_2$ compound with $T_c$ = 20 K, the <δ> = 0.55 mm/s at 5 K is observed [48]. This suggests that Ba(Fe$_{1-x}$Ru$_x$)$_2$As$_2$ can be regarded as a truly isovalent doped system because its electron density measured by Mössbauer spectroscopy is exactly the same as in BaFe$_2$As$_2$ even at high substitution level x.

The temperature evolution of spectral shift <δ>, shown in **Fig. 3,** represents only a typical second-order Doppler shift, $\delta_{SOD}$, dependence on temperature. This is a relativistic effect due to thermal motion of the absorbing nuclei, and hence it could be treated in terms of the Debye model for the lattice vibrations of Fe atoms. This effect is proportional to the mean-square velocity $\langle v^2 \rangle$ of the Mössbauer nuclei and can be expressed as: $\delta_{SOD} = -\langle v^2 \rangle E_\gamma / 2c^2$, where $E_\gamma$ is the energy of resonant γ-photons. In the Debye approximation it has the following form: $\delta_{SOD}(T) = -\frac{9k_B T}{2Mc}\left(\frac{T}{\Theta_D}\right)^3 \int_0^{\Theta_D/T} \frac{x^3 dx}{e^x - 1}$, where $k_B$ is the Boltzmann constant, *M* is the mass of Mössbauer nucleus, *c* is the speed of light, and $\Theta_D$ is the Debye temperature. Here, the parameter $\Theta_D$ should not be considered as universal quantity, but rather as an effective local variable that is specific for the detection method. Our experimental data show that the Debye temperature $\Theta_D$ is typical for a strongly bound metal-covalent system. It increases with phosphorus substitution level from 419(7) K for the parent compound, through 456(4) K for a superconductor with the highest $T_c$, to 470(5) K for over-doped compounds. It means that the Fe−As/P bonds become stiffer with increasing P content. The present experimental data evaluated with the EFG distribution and without magnetic interactions indicate that the spectral shifts <δ> for doping levels x = 0.31 (for clarity not shown in **Fig. 3**) and x = 0.33 almost overlap between 80 and 300 K. They are, however, not in accordance with the experimental data from the low temperature range, where contribution from magnetic interactions becomes significant and makes the spectra extremely complex. Thus, the resulting $\Theta_D$ = 464(7) K for x = 0.31 was estimated using the data from "high" temperature regime.

For comparison the Debye temperature of the Ba$_{0.6}$K$_{0.4}$Fe$_2$As$_2$ superconductor, $\Theta_D$ = 459(4) K, was calculated using the δ(*T*) dependence and partly unpublished data reported in Ref. [23] (see Supplemental Material at [49]). Similar values of the Debye temperatures for the optimally doped BaFe$_2$(As$_{0.67}$P$_{0.33}$)$_2$ and Ba$_{0.6}$K$_{0.4}$Fe$_2$As$_2$ superconductors strongly indicate that the dynamics of Fe in both compounds is very similar as it depends on the "inner" dynamics of nearly the same Fe−As/P (pnictogens) sheets. On the other hand, the values of $\Theta_D$ found here are much larger than $\Theta_D$ = 250 K reported for BaFe$_2$As$_2$ and derived from the specific heat measurements [50]. This discrepancy is most likely due to different weights of the phonon frequency distributions used to determine $\Theta_D$ from the specific heat data and the Mössbauer spectroscopy parameters.

The Mössbauer spectroscopy recoilless fraction *f* expressed by the Lamb-Mössbauer factor (the analog of the Debye–Waller factor in the coherent neutron and X-ray scattering) is another parameter sensitive to lattice dynamics. It can be given in terms of the γ-energy $E_\gamma$ and the mean local displacement of nucleus from its equilibrium position $f = \exp[-\langle x^2 \rangle E_\gamma^2 / (\hbar c)^2]$, where $\langle x^2 \rangle$ is the squared vibrational amplitude (the mean-squared displacement) in the direction of γ-ray propagation. The parameter $f$ is proportional to the spectral area and can be expressed in the Debye approximation as follows:

$$f(T) = \exp\left\{-\frac{3E_\gamma^2}{4Mc^2 k_B \Theta_D}\left[1 + 4\left(\frac{T}{\Theta_D}\right)^2 \int_0^{\Theta_D/T} \frac{xdx}{e^x - 1}\right]\right\}.$$

We should note, however, a difference between parameters $\Theta_D$ determined from temperature dependence of the Mössbauer recoilless fraction $f$ and temperature dependence of spectral shift $\langle\delta\rangle$. The former is always significantly smaller than the latter. In fact, the $\Theta_D$ determined from $f$ is fairly similar to $\Theta_D$ derived from specific heat data. This is mainly because both recoilless fraction and specific heat within the approach of the Debye theory are related to the mean-squared displacement $\langle x^2 \rangle$, which weights the phonon frequency distribution by $\omega^{-1}$, whereas $\delta_{SOD}$ is related to the mean-squared velocity $\langle v^2 \rangle$, which weights the lattice vibration frequency distribution by $\omega^{+1}$ [51].

In the present studies, the Mössbauer recoilless fraction $f$ is approximated by the relative spectral area (*RSA*), which is proportional to $f$ and defined as:

$$RSA = \left(\frac{1}{C}\right) \sum_{n=1}^{C} \frac{N_0 - N_n}{N_0},$$

where $C$ denotes the number of data channels for folded spectrum, $N_0$ is the average number of counts per channel far-off resonance, namely the baseline, and $N_n$ stands for the number of counts in the channel $n$. The *RSA* is especially useful in describing the measured $BaFe_2(As_{1-x}P_x)_2$ spectra due to multitude of the hyperfine interactions affecting their shape. Moreover, the parameter *RSA* can be directly and easily evaluated from measurements as a quantity independent of any physical model. The temperature evolutions of the relative spectral areas for under-doped and superconducting $BaFe_2(As_{1-x}P_x)_2$ are displayed in **Fig. 4**. Here, the *RSA*s are calculated for the Mössbauer spectra recorded versus increasing temperature in the non-interrupted series of measurements carried out in the same experimental conditions (radioactive source, geometry, velocity scale, number of data channels, background under the resonance line, and linear response range of the detector system). A slight increase of *RSA* at 130 K for x = 0.10 is observed due to an increase in the average recoilless fraction arising from the magneto-elastic effect, resulting in a hardening of lattice vibrations upon development of the spin-nematic phase with the incoherent magnetic order. This feature seems common for the iron–pnictogen bonds [52, 53]. On the other hand, the *RSA* for x = 0.31, 0.33, and 0.53 across superconducting transition is flat and without any irregularity. Thus, spectral parameters dependent on the lattice dynamics, e.g., recoilless fraction (approximated by *RSA*) and second-order Doppler shift (included in $\langle\delta\rangle$) are insensitive to the transition. Hence, the lattice dynamics seen by [57]Fe Mössbauer spectroscopy seems unaffected by a transition to the superconducting state.

*4.1 Mössbauer hyperfine parameters calculated by DFT methods*

In order to qualitatively describe modifications of the hyperfine parameters resulting from the substitution of As atoms by P atoms in the $BaFe_2(As_{1-x}P_x)_2$ system, we resort in the following to parameter-free first-principles calculations realized by the DFT method. In particular, we examine the electron density (isomer shift), the electric field gradient (quadrupole splitting), and the hyperfine magnetic field. Insets in **Fig. 5** show examples of the modeled BaFe$_2$(As$_{1-x}$P$_x$)$_2$ compositions at x = 0.31 and x = 0.5.

Substitution of As atoms with P atoms yields slight local distortions of the FeT$_4$ (T = As, P) tetrahedra, which come from small changes in the interatomic distances between Fe atoms and the surrounding As/P atoms. The Fe-As (2.29 Å) and Fe-P (2.18 Å) distances in ideal BaFe$_2$As$_2$ and BaFe$_2$P$_2$ compounds become modified by 0.01 – 0.02 Å.

Moreover, the valence charges of ions forming Fe$_2$T$_2$ (T = As, P) layers in BaFe$_2$(As$_{1-x}$P$_x$)$_2$ compositions undergo changes, as indicated by the electron topological analysis [54]. Ideal compositions (x = 0.0, 1.0) are characterized by the following effective valencies: Ba$^{1.20+}$Fe$_2^{0.11+}$As$_2^{0.71-}$, and Ba$^{1.20+}$Fe$_2^{0.15+}$P$_2^{0.75-}$. We note evident departure of Fe and T (T = As, P) valences from their formal values (Fe$^{2+}$, T$^{2-}$) due to the covalent bonding within the Fe$_2$T$_2$ (T = As, P) layers. Effective charges of Fe ions in BaFe$_2$(As$_{1-x}$P$_x$)$_2$ at x = 0.31 and x = 0.5 are collected in **Table 1**. One observes an increase in the effective valence charge of Fe with increased number of P atoms in its neighborhood. In the limiting cases (4As and 4P configurations) the effective valences of Fe correspond to those in ideal BaFe$_2$As$_2$ and BaFe$_2$P$_2$ systems.

**Table 1.** Effective valence charges of Fe ions in BaFe$_2$(As$_{1-x}$P$_x$)$_2$ with x = 0.31 and x = 0.5 determined from the electron topological analysis [54]. Values are averaged over all considered configurations.

| Substitution | Effective valence charges (in $e$) for respective configuration | | | | |
|---|---|---|---|---|---|
| x | 4As | 3As1P | 2As2P | 1As3P | 4P |
| 0.31 | 0.11+ | 0.12+ | 0.14+ | 0.15+ | 0.16+ |
| 0.50 | 0.11+ | 0.12+ | 0.14+ | 0.15+ | - |

The changes in local environment of Fe, which arise from appearance of substitutional P ions in its closest surrounding, are also reflected by the calculated isomer shifts, which decrease with increased number of incorporated P atoms and increased valency of Fe. The range of changes in the calculated isomer shifts for 4As, 3As1P, 2As2P, 1As3P, and 4P configurations in BaFe$_2$(As$_{1-x}$P$_x$)$_2$ at x = 0.0, 0.31, 0.50, 1.0 are given in **Table 2**. The calculated chemical isomer shifts IS does not include contribution from the second-order Doppler effect ($\delta_{SOD}$), which is of the order of -0.1 mm/s [32].

**Table 2.** The calculated isomer shifts of Fe residing in the 4As, 3As1P, 2As2P, 1As3P, and 4P environments in BaFe$_2$(As$_{1-x}$P$_x$)$_2$ with x = 0.0, 0.31, 0.5, 1.0. The isomer shifts are given versus bcc α-Fe. Values are averaged over all configurations considered.

| Substitution | Isomer shift (mm/s) for respective configuration | | | | |
|---|---|---|---|---|---|
| x | 4As | 3As1P | 2As2P | 1As3P | 4P |
| 0.0 | 0.373 | - | - | - | - |
| 0.31 | 0.354-0.361 | 0.330-0.340 | 0.304-0.326 | 0.284-0.298 | 0.261-0.275 |
| 0.50 | 0.371 | 0.338 | 0.322-0.324 | 0.300 | 0.275 |
| 1.0 | - | - | - | - | 0.202 |

In general, the quadrupole splittings in BaFe$_2$(As$_{1-x}$P$_x$)$_2$ with x = 0.31 show quite a broad distribution spanning the range between 0.14 and 0.58 mm/s. One can, however, distinguish two sub-ranges covering values 0.14 – 0.32 mm/s and 0.34 – 0.58 mm/s. The former one corresponds to 4As and 4P configurations, whereas the latter one is due to 3As1P, 2As2P, and 1As3P surroundings. In composition with x = 0.50, the Fe atoms in 4As and 4P environments exhibit quadrupole splittings of about 0.46 mm/s and 0 mm/s, respectively, while quadrupole splittings of Fe atoms in 1As3P, 2As2P, and 3As1P configurations range between 0.31 and 0.34 mm/s. For both compositions, the 2As2P and 3As1P tetrahedra remain more distorted than those of 1As3P, which gives rise to slightly larger quadrupole splitting of Fe atoms residing in 2As2P and 3As1P configurations in comparison with Fe atoms inside 1As3P configuration. One could expect a difference in quadrupole splittings and isomer shifts of Fe atoms located in somewhat different 2As2P surroundings, which are distinct from each other with respect to the positions of two As (two P) atoms. The As/P atoms forming 2As2P tetrahedrons may lie within the same or different layers. The difference in quadrupole splittings between such configurations is less than 0.05 mm/s, whereas more than an order of magnitude smaller difference (~0.002 mm/s) is observed for the respective isomer shifts. The average quadrupole splitting $<\Delta>$ has a significantly higher value for $x = 0.31$ than for $x = 0.5$ which agrees very well with the experimental data. The calculated average hyperfine magnetic field in $BaFe_2(As_{1-x}P_x)_2$ with $x = 0.31$ equals about 3.2 T as compared to that extracted from the spectrum measured for sample having the same composition (2.2 T). According to the experimentally determined phase diagram of $BaFe_2(As_{1-x}P_x)_2$, the composition with $x = 0.5$ is still superconducting and shows no magnetic interactions, as confirmed by the present experimental studies (no magnetic field at Fe nuclei) as well as theoretical calculations ($<B>$ less than 0.1 T).

The DFT calculated hyperfine parameters could be applied to simulate the Mössbauer spectra for substitution levels $x = 0.31$ (optimally doped $BaFe_2(As_{1-x}P_x)_2$ superconductor) and $x = 0.5$. Theoretical spectra are shown in **Fig. 5**. They correlate relatively well with the low-temperature experimental spectra of $x = 0.31$, 0.33, and 0.53 compositions, presented in **Figs. 1 and 2**, which do not display magnetic interactions or those with average magnetic field less than 2 T. The average shift $<IS>$ for $x = 0.31$ is somewhat larger than that for $x = 0.5$ due to increase in the *s*-electrons charge density with increased level of P-for-As substitution.

## 5. Conclusions

The $BaFe_2(As_{1-x}P_x)_2$ iron-based superconductors are formally considered as isovalent-doped system. We found that this material should be recognized as the hole-doped system when considering the change of the electron charge densities at iron nuclei reflected by isomer shifts of the $^{57}Fe$ Mössbauer spectra.

Under-doped $BaFe_2(As_{0.90}P_{0.10})_2$ exhibits reduced amplitude of incommensurate SDW as compared to the parent compound. It preserves the universality class of well separated two-dimensional magnetic planes with one-dimensional spins. The spin-nematic phase temperature region for $x = 0.10$ is characterized by the incoherent spin density order. The $BaFe_2(As_{0.69}P_{0.31})_2$ shows some coexistence of the weak magnetic order and superconductivity probably due to vicinity of the quantum critical point.

The present Mössbauer studies demonstrate that the average EFG and/or the charge density modulations are somewhat perturbed near $T_c$ for the $BaFe_2(As_{0.67}P_{0.33})_2$ and $BaFe_2(As_{0.47}P_{0.53})_2$ superconductors. This phenomenon is, however, not as spectacular as in $Ba_{0.6}K_{0.4}Fe_2As_2$ [23] and $SmFeAsO_{0.91}F_{0.09}$ [24] mainly because of a complexity of the Mössbauer spectra caused by phosphorous dopants in the Fe-As layers. The optimally doped system with $x = 0.33$ and the highest $T_c$ shows pronounced hump in $<\Delta>$ across superconducting transition, which reaches 0.02 mm/s between 12 K and 28 K. The jump in $<\Delta>$ is partly recovered below $T_c$, when the superconducting gap is fully developed. The change in $<\Delta>$ can be converted into the change in the EFG and amounts to about $10^{20}$ V/m$^2$ at $T_c$. Therefore, a distribution of the electrons responsible for the covalent Fe-As/P bonds is somewhat perturbed by the itinerant electrons forming Cooper pairs in this metallic system.

Increase in <Δ> at $T_c$ for the system with x = 0.33 is quite similar to an increase in the EFGW amplitude at $T_c$ for electron-doped $SmFeAsO_{0.91}F_{0.09}$, and it has opposite behavior to the change of the EFGW amplitude at $T_c$ for hole-doped $Ba_{0.6}K_{0.4}Fe_2As_2$.

The phosphorus substitution increases the Debye temperature due to hardening of lattice vibrations. The spectral shift <δ>, which includes temperature-dependent second-order Doppler shift $δ_{SOD}$, and the spectral area $RSA$ are not affected by the transition to superconducting state. This indicates that neither the average electron density at Fe nuclei nor the dynamic properties of Fe atoms in lattice are sensitive to the superconducting transition.

The complexity of the experimental Mössbauer spectra could be to some extent resolved by calculating from the first-principles parameters of the electric monopole, electric quadrupole, and magnetic dipole interactions, which determine the measured spectral patterns. Modifications of such parameters as isomer shift, quadrupole splitting, and hyperfine field at the resonant iron nuclei reflect changes in their immediate surrounding originating from the incorporation of P atoms into As-sublattice.

**Acknowledgments**

This work was supported by the National Science Centre of Poland under the grant 2018/29/N/ST3/00705.

The work at Tohoku University was in part supported by Grant-in-Aids for Scientific Research (No. JP19H05824) from the Ministry of Education, Culture, Sports, Science and Technology (MEXT), Japan.

The DFT Group (DL and UDW) acknowledges the European Regional Development Fund in the IT4Innovations national supercomputing center - path to exascale project No. CZ.02.1.01/0.0/0.0/16_013/0001791 within the Operational Programme Research, Development and Education Czech Science Foundations and the Interdisciplinary Center for Mathematical and Computational Modeling (ICM), Grants No. GA73-17 and No. GB70-12 Warsaw University.

This article is dedicated to the memory of Professor Krzysztof Piotr Ruebenbauer.

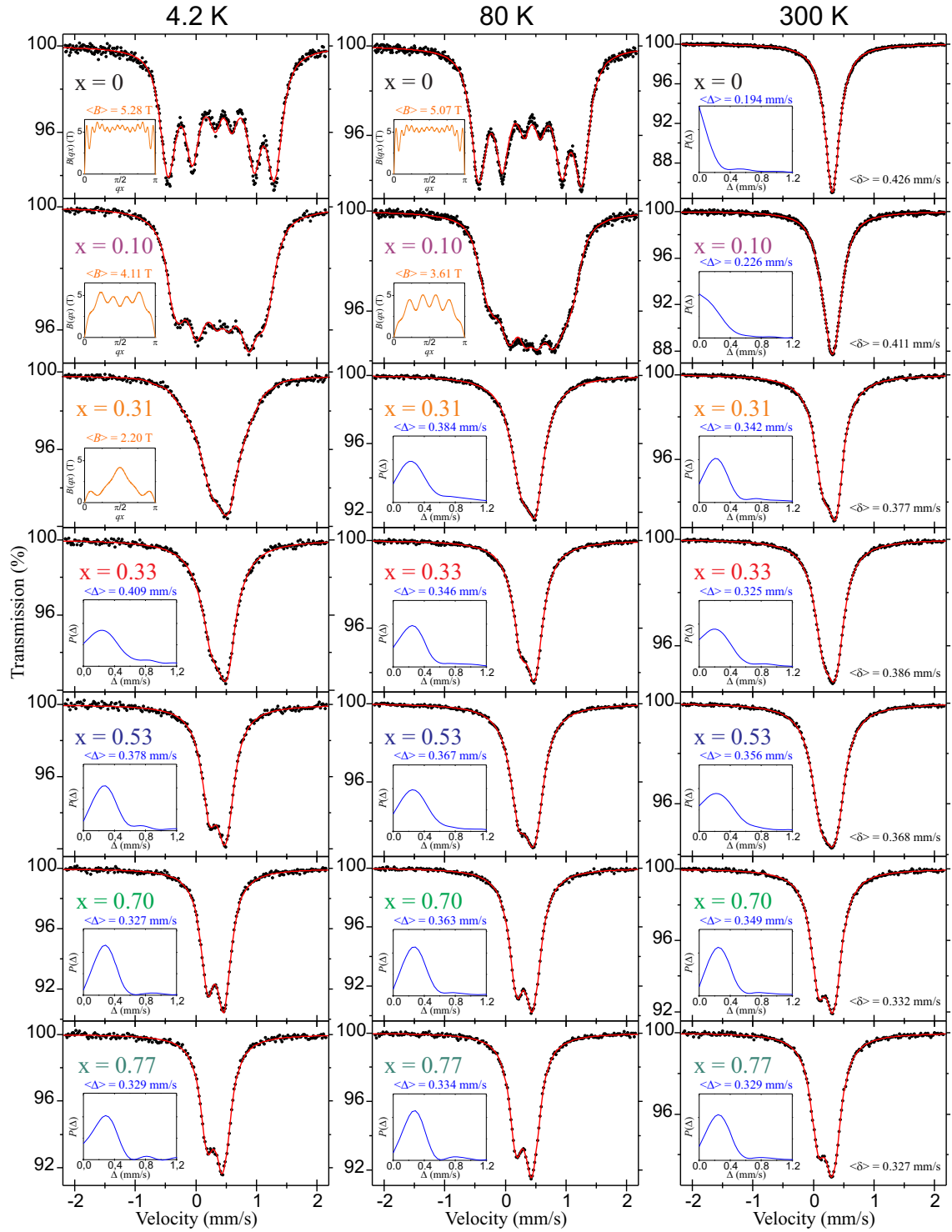

**Fig. 1** $^{57}$Fe Mössbauer spectra of BaFe$_2$(As$_{1-x}$P$_x$)$_2$ measured at 4.2 K, 80 K and 300 K. (Insets) Shapes of SDW with the average magnetic hyperfine field <$B$> and normalized distributions of the electric quadrupole splitting with the average value <$\Delta$>. The average center shifts <$\delta$> of spectra measured at 300 K are also included.

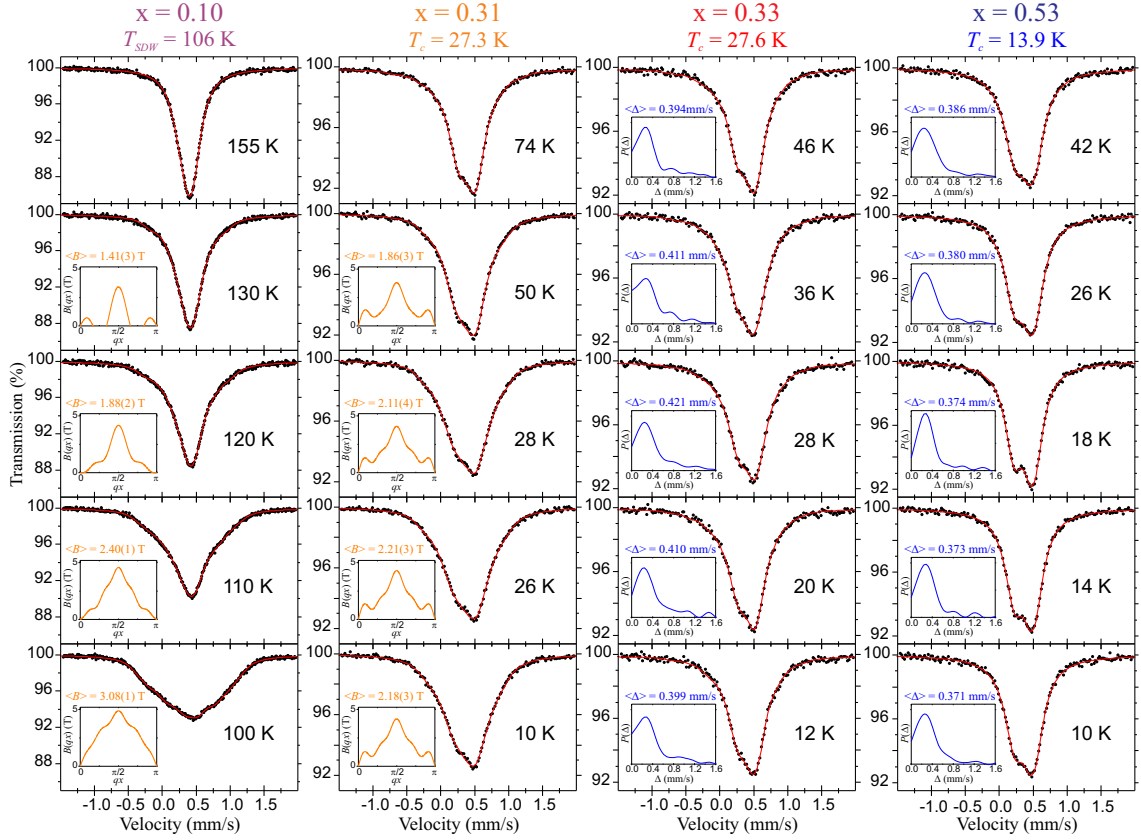

**Fig. 2** Selected $^{57}$Fe Mössbauer spectra of BaFe$_2$(As$_{1-x}$P$_x$)$_2$ versus temperature across the spin-nematic phase region (for x = 0.10) and across transition to the superconducting state (for x = 0.31, 0.33, and 0.53). Spectra measured for particular compounds at the same conditions during non-interrupted series are shown for comparison on the same scale. (Insets) Shapes of SDW with the average magnetic hyperfine field <B> and the distributions of the electric quadrupole splitting with the average value <Δ>. The SDW shapes should be interpreted as the incoherent spin density wavelets for the spin-nematic region. The temperature of the coherent SDW magnetic order $T_{SDW}$ and the critical temperatures of superconducting transitions $T_c$ are indicated in the headline.

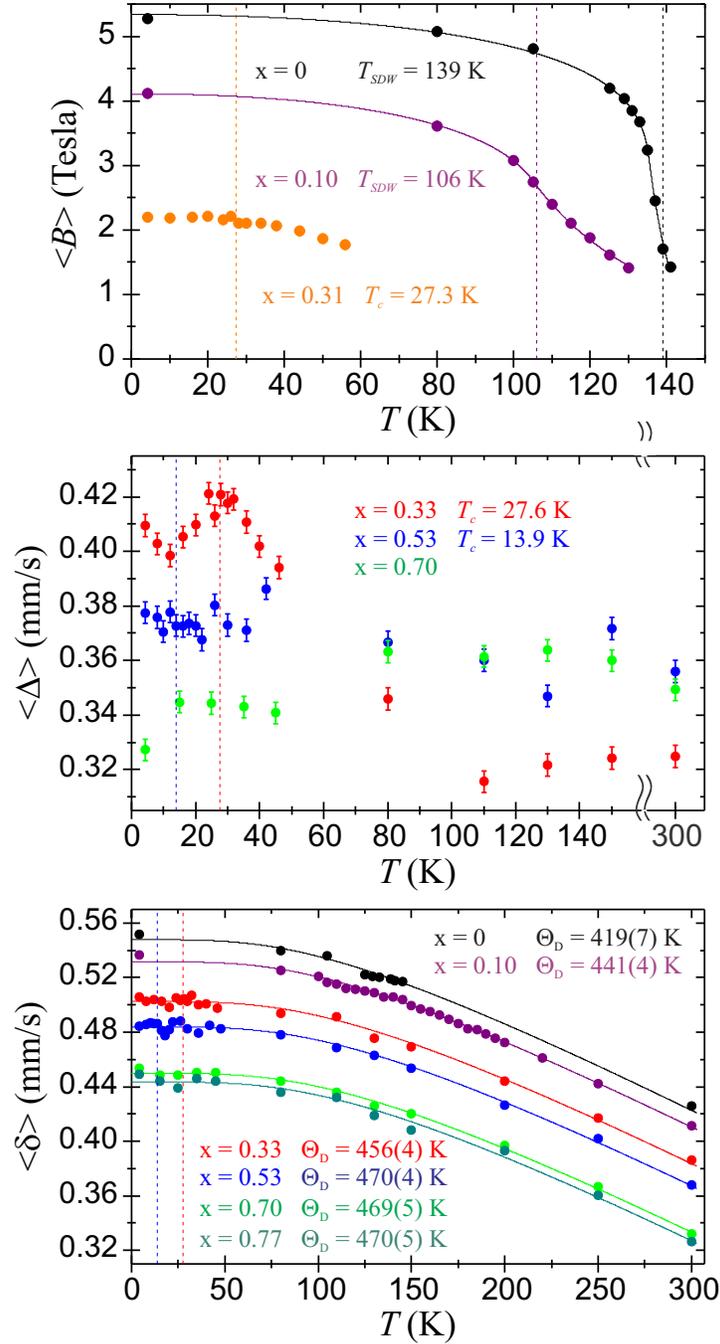

**Fig. 3** Temperature dependencies of the Mössbauer spectroscopy parameters (the average magnetic hyperfine field of SDW $\langle B \rangle$, the average electric quadrupole splitting $\langle \Delta \rangle$, and the average spectral center shift $\langle \delta \rangle$) for $BaFe_2(As_{1-x}P_x)_2$ compounds with various phosphorus substitution level x. The coherent SDW magnetic order temperatures $T_{SDW}$, the critical temperatures of superconducting transitions $T_c$, and the Debye temperatures $\Theta_D$ are shown for different phosphorus substitution level x. Vertical dashed lines denote temperatures $T_{SDW}$ and $T_c$. The solid lines represent the best-fit to experimental data using the Debye model for $\langle \delta \rangle(T)$ and the model described in Ref. [21] for $\langle B \rangle(T)$. Typical errors for $\langle B \rangle$ and $\langle \delta \rangle$ are 0.03 T and 0.003 mm/s, respectively.

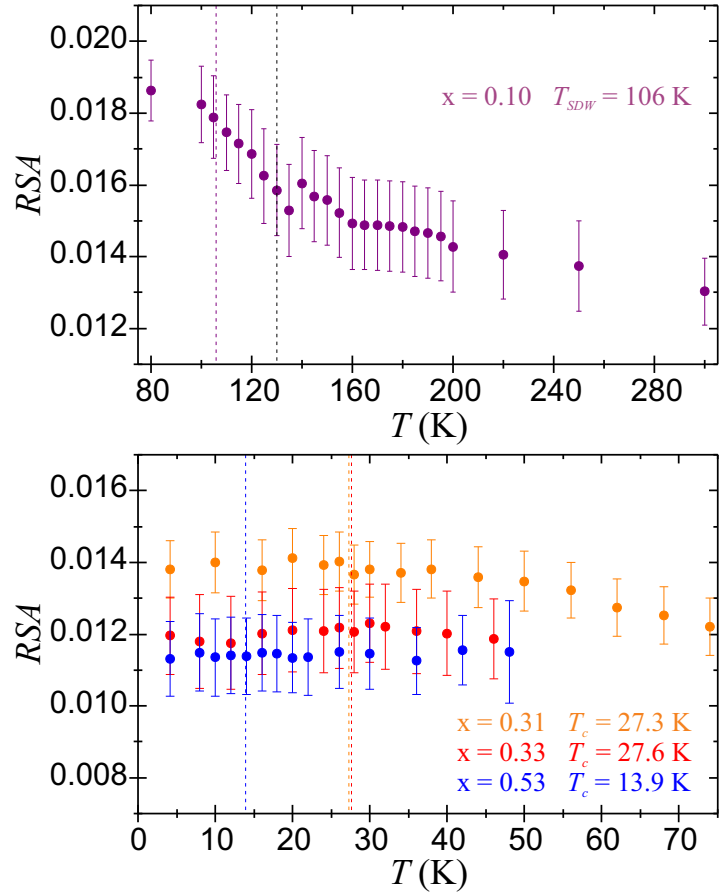

**Fig. 4** Temperature dependencies of the relative spectral areas (*RSA*) for under-doped and superconducting BaFe$_2$(As$_{1-x}$P$_x$)$_2$. The vertical dashed lines denote temperatures $T_{SDW}$ and $T_c$. The black dashed line at 130 K indicates the spin-nematic order temperature [7] for x = 0.10. Each *RSA* is derived from series of measurements performed at constant conditions (see text for details). Note: despite the identical weights of samples the sets of *RSA* may not be comparable between each other due to possible differences in the resonant thicknesses between particular absorbers.

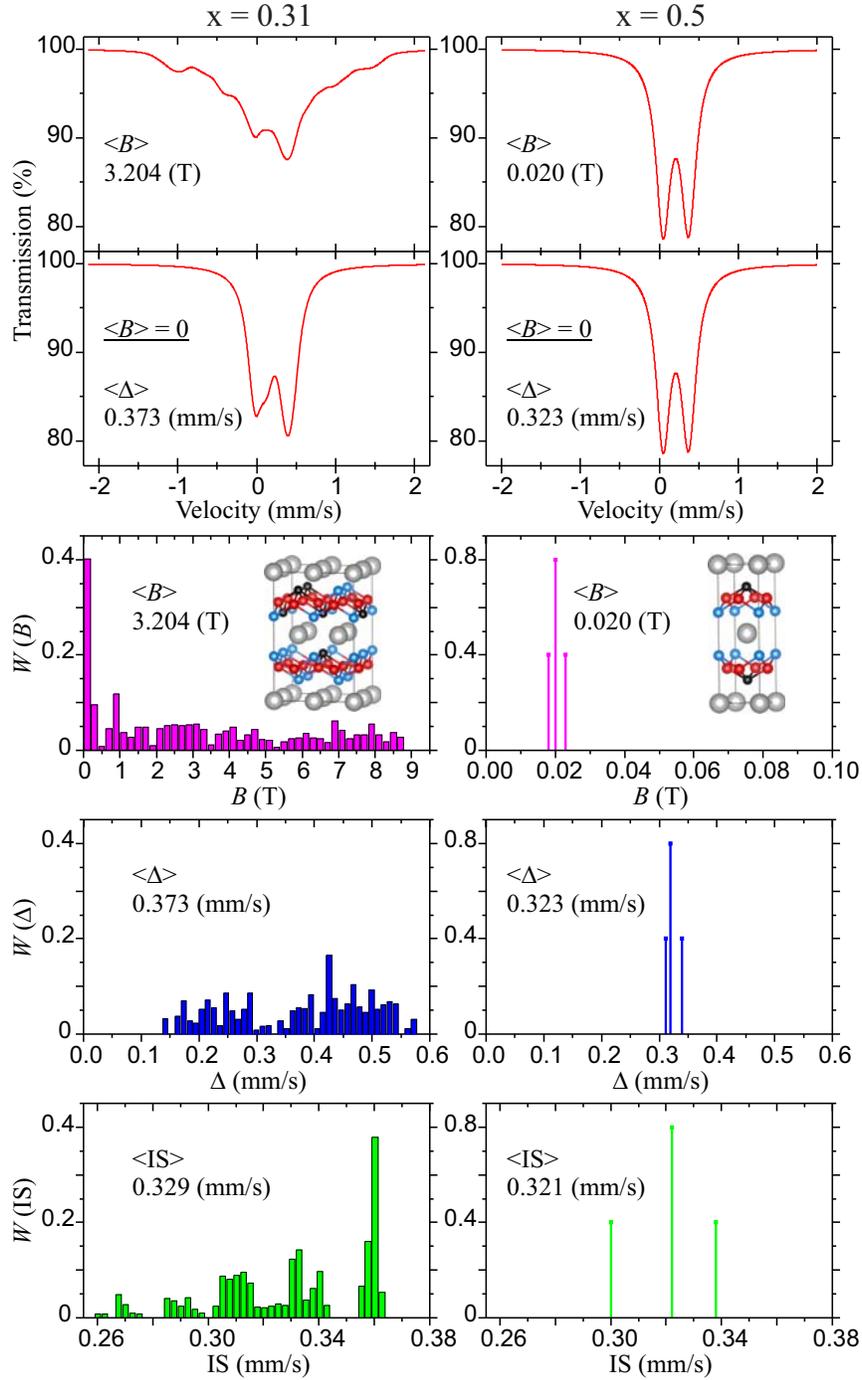

**Fig. 5** Simulated $^{57}$Fe Mössbauer spectra for BaFe$_2$(As$_{1-x}$P$_x$)$_2$ with x = 0.31 and x = 0.5, and calculated distributions of the magnetic hyperfine field $B$, the electric quadrupole splitting $\Delta$, and the chemical isomer shift IS versus α-Fe ($\delta_{SOD}$ is neglected). The numbers indicate respective average values. (Insets) Schematic examples of simulated configurations with gray, red, blue, and black balls denoting Ba, Fe, As, and P atoms, respectively.

Supplemental Material

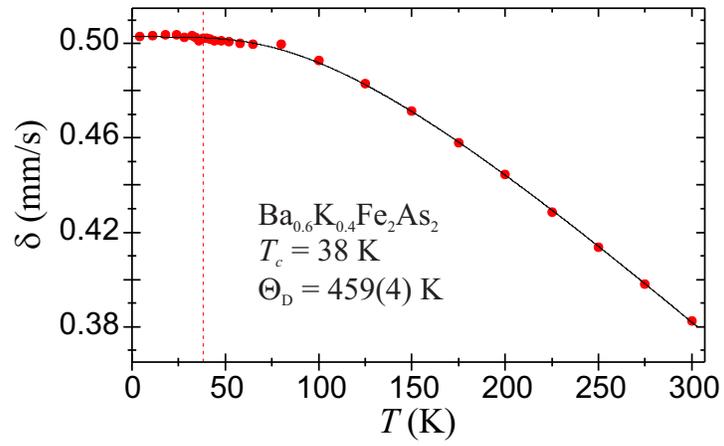

**Fig. S1** The Mössbauer spectra center shift δ versus temperature for optimally hole-doped $Ba_{0.6}K_{0.4}Fe_2As_2$ superconductor. The vertical dashed line denotes the critical temperature $T_c = 38$ K. The solid line represents the best-fit to experimental data according to the Debye model with resulting Debye temperature $\Theta_D = 459$ K.